\documentclass[12pt,onecolumn,oneside,a4paper,peerreview]{IEEEtran}
\usepackage[T1]{fontenc}
\usepackage{graphicx,latexsym,subfigure,amsmath,epsfig,latexsym,subfigure,amsmath,amssymb}
\usepackage[noadjust]{cite}
\usepackage{algorithm,amsthm}
\usepackage{cases}
\usepackage{color}
\usepackage{bm}
\usepackage{url}
\usepackage{multirow}
\usepackage{amssymb}
\usepackage{cases}
\usepackage{setspace}
\usepackage{algorithmic}
\usepackage{textcomp}
\usepackage{amsmath}
\usepackage{bm}
\usepackage{algorithm}
\usepackage{algorithmic}
\begin{document}
\title{PrecoderNet: Hybrid Beamforming for Millimeter Wave Systems with Deep Reinforcement Learning}
\author{
\begin{minipage}{0.98\columnwidth}
\vspace*{0.8cm}
\begin{center}
\IEEEauthorblockN{Qisheng Wang\IEEEauthorrefmark{1},
              Keming Feng\IEEEauthorrefmark{1},
              Xiao Li\IEEEauthorrefmark{1},
              and Shi Jin\IEEEauthorrefmark{1}\\
              %and Shi Jin\authorrefmark{1}\\
\vspace*{0.2cm} \small{
\IEEEauthorblockA{\IEEEauthorrefmark{1}National Mobile Communications Research Laboratory,
Southeast University, Nanjing 210096, China} } }
\end{center}
\end{minipage}
} \IEEEaftertitletext{\vspace{-0.75\baselineskip}}
% make the title area
\maketitle \vspace*{1.2cm}

%-----------------------------------------------------------------------------------------
% Abstract
%-----------------------------------------------------------------------------------------
\begin{abstract}
In this letter, we investigate the hybrid beamforming for millimeter wave massive multiple-input multiple-output (MIMO) system based on deep reinforcement learning (DRL). Imperfect channel state information (CSI) is assumed to be available at the base station (BS). To achieve high spectral efficiency with low time consumption, we propose a novel DRL-based method called PrecoderNet to design the digital precoder and analog combiner. The DRL agent takes the digital beamformer and analog combiner of the previous learning iteration as state, and these matrices of current learning iteration as action. Simulation results demonstrate that the PrecoderNet performs well in spectral efficiency, bit error rate (BER), as well as time consumption, and is robust to the CSI imperfection.
\end{abstract}

\begin{IEEEkeywords}
MmWave, hybrid beamforming, deep reinforcement learning, MMSE.
\end{IEEEkeywords}

\section{Introduction}
Millimeter wave (mmWave) communications have been considered as a potential technique to solve the frequency resource shortage problem \cite{niu2015survey}. Combined with massive multiple-input multiple-output (MIMO), sufficient antenna array gain can be provided to overcome the high penetration loss of mmWave signal. However, the design of its beamforming matrix is constrained by the expensive mmWave radio-frequency (RF) chains. Traditional full-digital beamformer needs to connect a RF chain for each antenna element, and thus imposes intolerant power consumption and hardware cost.

To solve this problem, a hybrid beamforming (HBF) architecture was proposed \cite{el2014spatially}. It replaces the full digital beamformer with a low-dimensional digital precoder followed by a high-dimensional analog precoder. To obtain feasible hybrid beamformers and combiners, some numerical algorithms, such as orthogonal matching pursuit (OMP) \cite{el2014spatially}, Karush-Kuhn-Tucker (KKT) based \cite{sohrabi2016hybrid}, and manifold optimization (MO) based \cite{yu2016alternating} algorithms, were proposed. To achieve near-optimal performance while reducing the long time consumption incurred by the conventional numerical algorithms, some deep learning (DL) based algorithms were proposed\cite{elbir2019deep,elbir2019cnn}. However, the DL method requires large amount of training data in advance. In some cases, the training data itself is very difficult to obtain. When the transmission environment changes, new training data is needed and the network needs to be retrained.

Recently, deep reinforcement learning (DRL) has attracted more and more attention \cite{gu2017deep,lillicrap2015continuous} due to its powerful ability to deal with non-convex problems. The DRL agent can use few shots to effectively learn the optimal behaviour policy to handle complex problems. Moveover, compared with DL \cite{xia2019deep,Li_Access}, it utilizes the samples generated previously to train the agent and does not need large amount data for off-line training.

Motivated by the above analysis, in this letter, we employ DRL to investigate the hybrid beamforming for mmWave systems. Assuming imperfect channel state information (CSI) at base station (BS), we develop a deep deterministic policy gradient (DDPG) \cite{lillicrap2015continuous} based algorithm called PrecoderNet to maximize an average rate upper bound. The digital beamformer and analog combiner of the previous learning iteration are taken as state, while these matrices of current learning iteration are selected as action. The average rate upper bound is regarded as reward. Simulations show that considerable performance in spectral efficiency, bit error rate (BER), and robustness can be achieved by the proposed algorithm with low time consumption.

\section{System model }
Consider a mmWave massive MIMO single-user downlink system which employs $N_t$ antennas connected with $N^t_{RF}$ RF chains at the base station (BS). The user has $N_r$ antennas connected with $N^{r}_{RF}$ RF chains, its received signal is \cite{el2014spatially}
\begin{equation}\label{eq1}
  \mathbf{y}=\mathbf{W}^H_{BB}\mathbf{W}^H_{RF}\mathbf{H}\mathbf{V}_{RF}\mathbf{V}_{BB}\mathbf{x}+\mathbf{W}^H_{BB}\mathbf{W}^H_{RF}\mathbf{n},
\end{equation}
where $\mathbf{y}\in\mathbb{C}^{N_s\times 1}$ and $\mathbf{x}\in\mathbb{C}^{N_s\times 1}$ are the received and transmitted signals, $\mathbb{E}[\mathbf{x}\mathbf{x}^H]=\mathbf{I}_{N_s}$, $\mathbf{n}\sim\mathcal{CN}(0, \sigma^2_{n}\mathbf{I}_{N_r})$ is the noise, $\mathbf{H}\in \mathbb{C}^{N_r\times N_t}$ is the channel matrix, $\mathbf{V}_{BB}\in \mathbb{C}^{N^{t}_{RF}\times N_s}$ and $\mathbf{W}_{BB}\in \mathbb{C}^{N^{r}_{RF}\times N_s}$ are the digital beamformer and combiner, $\mathbf{V}_{RF}\in \mathbb{C}^{ N_t\times N^{t}_{RF}}$ and $\mathbf{W}_{RF}\in \mathbb{C}^{ N_r\times N^{ r}_{RF}}$ are the analog beamformer and combiner satisfying $|\mathbf{V}_{RF}(i,j)|=|\mathbf{W}_{RF}(i,j)|=1$, $\mathbf{A}(i,j)$ represents the $(i, j)$-th element of matrix $\mathbf{A}$, $\mathbf{V}_{BB}$ and $\mathbf{V}_{RF}$ satisfy the constraint $\mathrm{Tr}\Big(\mathbf{V}_{RF}\mathbf{V}_{BB}\mathbf{V}_{BB}^H\mathbf{V}_{RF}^H\Big)\leq P$, $P$ is the maximum transmit power, and $(\cdot)^H$ denotes conjugate transpose.

Assuming that both the BS and user are equipped with uniform linear array (ULA), the well-known geometric Saleh-Valenzuela channel model \cite{raghavan2010sublinear} is adopted, i.e.,
\begin{equation}\label{eq2}
  \mathbf{H}=\sqrt{\frac{N_tN_r}{N_{cl}N_{ray}}}\sum_{i=1}^{N_{cl}}\sum_{j=1}^{N_{ray}}\alpha _{ij}\mathbf{f}_r(\varphi^r_{ij})\mathbf{f}^H_t(\varphi^t_{ij}),
\end{equation}
where $N_{cl}$ is the number of scattering clusters, each cluster contains $N_{ray}$ scattering rays, $\alpha_{ij}\sim \mathbb{CN}(0,\sigma^2_{\alpha,i})$ is the complex path gain of the $j$-th ray in the $i$-th cluster, $\mathbf{f}_r(\varphi^r_{ij})$ and $\mathbf{f}_t(\varphi^t_{ij})$ are the normalized receiver and transmitter array response, $\varphi^r_{ij}$ and $\varphi^t_{ij}$ are the angle of arrival (AoA) and angle of departure (AoD). For a ULA with $N$ antennas and antenna spacing $d$, its array response can be expressed as
\begin{equation}\label{eq3}
  \mathbf{f}(\varphi)=\frac{1}{N}[1,e^{j\frac{2\pi d}{\lambda}sin(\varphi )},...,e^{j(N-1)\frac{2\pi d}{\lambda}sin(\varphi )}]^T,
\end{equation}
where $\varphi$ is the AoA/AoD and $\lambda$ is the carrier wavelength. We assume that the channel matrix obtained at BS, denoted as $\tilde{\mathbf{H}}$, related to the actual channel matrix $\mathbf{H}$ as follows
\begin{equation}\label{eq_h_es}
\mathbf{H}=\sqrt{1-\beta^2}\tilde{\mathbf{H}}+\beta\Delta\mathbf{H},
\end{equation}
where $\Delta\mathbf{H}$ representing the channel error is a random matrix with i.i.d. $\mathbb{CN}(0,1)$ elements, and $\beta\in[0,1]$ indicates the level of imperfection, which is known at the BS.

The spectral efficiency under this system model is \cite{sohrabi2016hybrid}
\begin{equation}\label{eq4}
R=\log_{2}\det\Big(\mathbf{I}_{N_s}+\mathbf{C}_{n}^{-1}\mathbf{W}^H\mathbf{H}\mathbf{V}\mathbf{V}^H\mathbf{H}^H\mathbf{W}\Big),
\end{equation}
where $\mathbf{C}_{n} = \sigma_{n}^{2}\mathbf{W}^H_{BB}\mathbf{W}^H_{RF}\mathbf{W}_{RF}\mathbf{W}_{BB}$, $\mathbf{V}=\mathbf{V}_{RF}\mathbf{V}_{BB}$ and $\mathbf{W}=\mathbf{W}_{RF}\mathbf{W}_{BB}$. Assuming perfect CSI at the BS, some sub-optimal numerical algorithms that decouple the beamformer and combiner design, and solve them in a sequential manner, i.e., OMP, KKT-based, and MO algorithm, were proposed to maximize (\ref{eq4}). However, the OMP method requires high transmit signal-to-noise ratio (SNR) and good sparsity of the digital precoding matrix, which might not be satisfied in practice. The KKT-based and MO-based algorithm are extremely time-consuming.
Moreover, directly maximizing (\ref{eq4}) with imperfect CSI at BS will lead to performance degradation. An effective approach under imperfect CSI is to maximize the average rate $\mathbb{E}_{\Delta \mathbf{H}}[R]$. However, it is difficult to obtain analytical expression for $\mathbb{E}_{\Delta \mathbf{H}}[R]$. Thus, we derive an analytical upper bound of $\mathbb{E}_{\Delta\mathbf{H}}[R]$, and try to design the hybrid precoders and combiners maximizing this upper bound.
Substituting (\ref{eq_h_es}) into (\ref{eq4}), and applying the Jensen's inequality, i.e., $\mathbb{E}[\log_2\det(\cdot)]\leq\log_2\det(\mathbb{E}[\cdot])$, $\mathbb{E}_{\Delta \mathbf{H}}[R]$ can be upper bounded as
\begin{equation}\label{eq_avgR}
\begin{aligned}
\mathbb{E}_{\Delta\mathbf{H}}[R]\leq\log_2\det&\Big[\Big(1+\frac{\beta^2P}{\sigma_n^2}\Big)\mathbf{I}_{N_s}+(1-\beta^2)\mathbf{C}_{n}^{-1}\mathbf{W}^H\tilde{\mathbf{H}}\mathbf{V}\mathbf{V}^H\tilde{\mathbf{H}}^H\mathbf{W}\Big]\overset{\triangle}=\bar{R}.
\end{aligned}
\end{equation}
%where inequality (a) is based on the Jensen inequality, i.e., $\mathbb{E}[\log_2\det(\cdot)]\leq\log_2\det(\mathbb{E}[\cdot])$, and equality (b) is based on the zero mean of random variable $\Delta\mathbf{H}$, i.e., $\mathbb{E}[\Delta\mathbf{H}]=0$.
%In the rest of this paper, we aim to maximize $\bar{R}$ to solve the hybrid beamforming problem.
Inspired by its ability to provide near-global optimal solution for non-concave problem \cite{mnih2015human} without large training dataset, in the rest of this letter, we try to solve this HBF problem utilizing the DRL approach DDPG to maximize $\bar{R}$ with lower time consumption.

\section{DRL-based HBF Design}
In this section, we first introduce the basic concepts of DRL, and then propose a so-called PrecoderNet to design the HBF matrixes utilizing DRL algorithm.

\subsection{Basic knowledge of DRL}
The DRL algorithm is composed of an agent interacting with the environment. The interactions between them can be denoted by a quintuple $<\mathcal{S},\mathcal{A},\emph{r},\emph{$\gamma$},\mathcal{P}>$, where $\mathcal{S}$ and $\mathcal{A}$ are the state and action space, \emph{r} is the reward fed back from environment to assess the selected action under current state, $\mathcal{P}$ is the behaviour policy depending on which the agent selects an action $a\in\mathcal{A}$ under a certain state $s\in\mathcal{S}$, i.e., $a=\mathcal{P}(s)$, and $\gamma\in(0,1]$ is a discount factor to ensure convergency within finite steps.

DDPG is a DRL algorithm that can handle the non-concave problem with continuous action space. A DDPG agent consists of an actor network $A$ and a critic network $C$ which are neural networks. At the $t$-th learning iteration, the actor network $A$ parameterized by $\bm{\theta}^A$ generates action $a^{(t)}$ under the state $s^{(t)}$, i.e., $a^{(t)}=A(s^{(t)}|\bm{\theta}^A)$, obtains reward $r^{(t)}$ from environment, and steps forward to the next state $s^{(t+1)}$. The critic network $C$ parameterized by $\bm{\theta}^C$ is a Q-network whose output is trained to approximate the Q-function
\begin{equation}\label{eq6}
\begin{aligned}
Q(s^{(t)},a^{(t)})= & \alpha Q(s^{(t)},a^{(t)})+(1-\alpha)[r^{(t)}+\gamma\mathop{\max}\limits_{a'\in\mathcal{A}}Q(s^{(t+1)},a')],
\end{aligned}
\end{equation}
where $\alpha$ is the learning rate. Its output $Q^C(s^{(t)},a^{(t)}|\bm{\theta}^C)$ is used to evaluate the state-action pair $(s^{(t)},a^{(t)})$. Then the actor and critic networks are updated to maximize the output of the critic network by gradient descent after randomly sampling in an experience replay $D$. In this way, the agent can learn the optimal action policy.

\subsection{DRL-based Hybrid Beamforming}
\begin{figure}[!ht]
\centering
\includegraphics[width=4.5in]{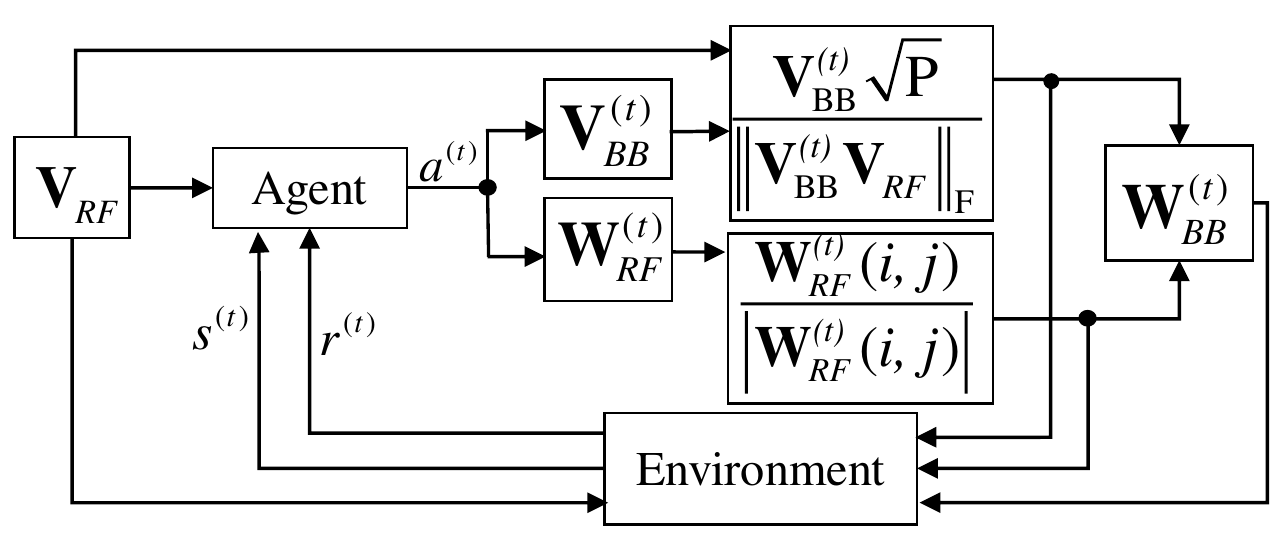}% 1\linewidth
\caption{Flow chart of PrecoderNet.}\label{fig2}
\end{figure}

In this section, a DDPG-based architecture referred to as PrecoderNet, shown in Fig. \ref{fig2}, is proposed to design the precoders and combiniers based on the upper bound (\ref{eq_avgR}). The environment is the whole transmission system seen at BS. In the left-most module, we obtain an applicable analog beamoformer $\mathbf{V}_{RF}$. Then, a DDPG agent is used to learn $\mathbf{V}_{BB}$ and $\mathbf{W}_{RF}$ simultaneously based on (\ref{eq_avgR}), and the corresponding $\mathbf{W}_{BB}$ is obtained based on the MMSE criterion in each learning iteration\footnote{Note that we can alternatively learn $\mathbf{W}_{BB}$ by the agent and obtain $\mathbf{W}_{RF}$ through MMSE criterion. However, the MMSE algorithm requires matrix inversion, whose calculation complexity is proportional to the matrix dimension. Due to the high-dimension of $\mathbf{W}_{RF}$, the computation complexity of obtaining $\mathbf{W}_{RF}$ is much higher than obtaining $\mathbf{W}_{BB}$ through MMSE criterion. Thus, we use the agent to learn $\mathbf{W}_{RF}$ so as to simplify the problem.}. At each learning iteration $t$, the agent observes the state $s^{(t)}$ from the environment and outputs action $a^{(t)}$, which consists of the digital beamformer $\mathbf{V}_{BB}^{(t)}$ and the analog combiner $\mathbf{W}_{RF}^{(t)}$. The two modules followed by $\mathbf{V}_{BB}^{(t)}$ and $\mathbf{W}_{RF}^{(t)}$ are used to satisfy the transmit power and constant modulus constraints. After that, the digital combiner $\mathbf{W}_{BB}^{(t)}$ is calculated based on MMSE criterion with the obtained $\mathbf{V}_{RF}$, $\mathbf{V}_{BB}^{(t)}$ and $\mathbf{W}_{RF}^{(t)}$. Then, the environment calculates the reward $r^{(t)}$ based on the obtained $\mathbf{V}_{RF}$, $\mathbf{V}_{BB}^{(t)}$, $\mathbf{W}_{RF}^{(t)}$, and $\mathbf{W}_{BB}^{(t)}$, and forwards it to the agent for its update. The whole process is further given in detail as follows.

It has been found in \cite{el2014spatially} that the optimal analog precoder can be chosen from the array responses of the transmitter. However, it is difficult to obtain precise AoA/AoD in practice. Therefore, the MO-based approach \cite{yu2016alternating}, which does not require the estimation of antenna array responses, is adopted in this letter to calculate feasible analog beamformer $\mathbf{V}_{RF}$. In this step, the CSI at BS is treated as perfect CSI to obtain $\mathbf{V}_{RF}$. The approach proposed in \cite{yu2016alternating} utilizes the MO algorithm to calculate all the precoders and combiners, while we only use MO to obtain $\mathbf{V}_{RF}$ to reduce the time consumption.

The design goal of $\mathbf{V}_{BB}$ and $\mathbf{W}_{RF}$ is to maximize $\bar{R}$. Therefore, at learning iteration \emph{t}, the corresponding upper bound $\bar{R}^{(t)}$ obtained by (\ref{eq_avgR}) is taken as the reward $r^{(t)}$. The combination of digital beamformer $\mathbf{V}_{BB}^{(t)}$ and analog combiner $\mathbf{W}_{RF}^{(t)}$ is the action $a^{(t)}$, i.e., $a^{(t)}=\{\mathbf{V}_{BB}^{(t)},\mathbf{W}_{RF}^{(t)}\}$, and the combination of them in the previous learning iteration is taken as the state $s^{(t)}$, i.e., $s^{(t)}=\{\mathbf{V}_{BB}^{(t-1)},\mathbf{W}_{RF}^{(t-1)}\}$. Based on MMSE criterion \cite{sohrabi2016hybrid}, %and the imperfection level of CSI obtained at the BS,
$\mathbf{W}_{BB}^{(t)}$ can be given as
\begin{equation}\label{eq7}
\mathbf{W}_{BB}^{(t)}=\sqrt{1-\beta^2}\Big(\mathbf{W}_{RF}^{(t)H}\mathbf{\Psi}^{(t)}\mathbf{W}_{RF}^{(t)}\Big)^{-1}\mathbf{W}_{RF}^{H}\tilde{\mathbf{H}}\mathbf{V}^{(t)},
\end{equation}
where $\mathbf{\Psi}^{(t)}=(1-\beta^2)\tilde{\mathbf{H}}\mathbf{V}^{(t)}(\mathbf{V}^{(t)})^H\tilde{\mathbf{H}}^H+\Big(\beta^2P+\sigma_n^2\Big)\mathbf{I}_{N_r}$,
and $\mathbf{V}^{(t)}=\mathbf{V}_{RF}\mathbf{V}_{BB}^{(t)}$.
\begin{figure}[!ht]
\centering
\includegraphics[width=4.5in]{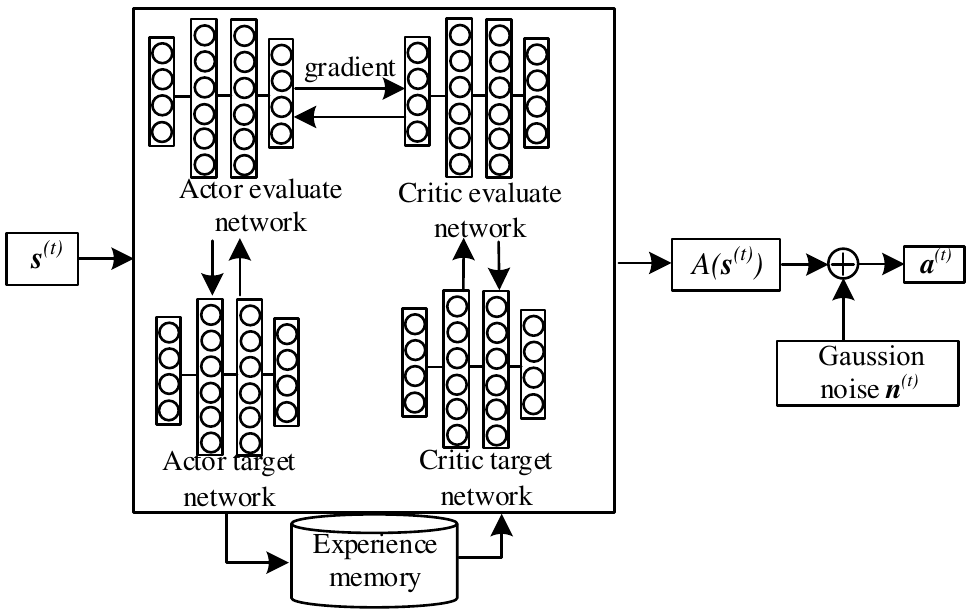}% 1\linewidth
\caption{Network architecture of hybrid beamforming agent.}\label{fig3}
\end{figure}

The framework of the agent is shown in Fig. \ref{fig3}. At learning iteration $t$, the agent obtains the corresponding state $s^{(t)}$ and reshapes the matrix $\mathbf{V}^{(t-1)}_{BB}$ and $\mathbf{W}^{(t-1)}_{RF}$ into the following vector, which is taken as the the input of the network, i.e.,
\begin{equation}\label{eq8}
\begin{aligned}
\mathbf{s}^{(t)}=\Big[ & \mathrm{Re}(\mathrm{vec}(\mathbf{V}^{(t-1)}_{BB}))^T,\mathrm{Im}(\mathrm{vec}(\mathbf{V}^{(t-1)}_{BB}))^T,\mathrm{Re}(\mathrm{vec}(\mathbf{W}^{(t-1)}_{RF}))^T,\mathrm{Im}(\mathrm{vec}(\mathbf{W}^{(t-1)}_{RF}))^T \Big]^T,
\end{aligned}
\end{equation}
where $\mathrm{Re}(\cdot)$ and $\mathrm{Im}(\cdot)$ denote the real and imaginary part of the input. The actor network $A$ outputs a vector $A(\mathbf{s}^{(t)})\in\mathbb{R}^{K\times1}$ and adds a noise vector $\mathbf{n}^{(t)}\in\mathbb{R}^{K\times1}$ to $A(\mathbf{s}^{(t)})$, where $K=2(N_{RF}^tN_s+N_rN_{RF}^r)$, for exploration to obtain the vector $\mathbf{a}^{(t)}\in\mathbb{R}^{K\times1}$, i.e., $\mathbf{a}^{(t)}=A(\mathbf{s}^{(t)})+\mathbf{n}^{(t)}$. After that, $\mathbf{a}^{(t)}$ is reorganized into $\mathbf{V}_{BB}^{(t)}$ and $\mathbf{W}_{RF}^{(t)}$ so that
\begin{equation}\label{eq9}
\begin{aligned}
\mathbf{a}^{(t)} = \Big[ & \mathrm{Re}(\mathrm{vec}(\mathbf{V}^{(t)}_{BB}))^T,\mathrm{Im}(\mathrm{vec}(\mathbf{V}^{(t)}_{BB}))^T,\mathrm{Re}(\mathrm{vec}(\mathbf{W}^{(t)}_{RF}))^T,\mathrm{Im}(\mathrm{vec}(\mathbf{W}^{(t)}_{RF}))^T \Big]^T.
\end{aligned}
\end{equation}
Then, $\mathbf{W}_{RF}^{(t)}$ and $\mathbf{V}^{(t)}_{BB}$ are normalized to satisfy the constant modulus and total power constraints respectively. The agent obtains the next state $s^{(t+1)}=\{\mathbf{V}_{BB}^{(t)},\mathbf{W}_{RF}^{(t)}\}$ and the reward $r^{(t)}=\bar{R}^{(t)}$. Moreover, the tuple $[\mathbf{s}^{(t)},\mathbf{a}^{(t)},r^{(t)},\mathbf{s}^{(t+1)}]$ is stored into an replay buffer $D$ with capacity $N_D$. The critic network $C$ evaluates the state-action pair by sampling a $N$-size minibatch prior experience $\mathcal{E} = \{e_1,...,e_N\}$, where $e_i=[\mathbf{s}^{(t_i)},\mathbf{a}^{(t_i)},r^{(t_i)},\mathbf{s}^{(t_i+1)}]$, ${1\leq i\leq N}$, from replay buffer $D$ and uses the mean of the previous experiences to calculate the approximated value $Q^{C}(\mathbf{s}^{(t_i)},\mathbf{a}^{(t_i)})$. This value is trained to approximate the target obtained in (6), and the loss function is given as
\begin{equation}\label{eq10}
L(\bm{\theta}^{C})=\frac{1}{N} \sum_{i=1}^{N}(Q^{C}(\mathbf{s}^{(t_i)},\mathbf{a}^{(t_i)})-y^{(t_i)})^{2}.
\end{equation}
where the target $y^{(t_i)}$ is obtained by (\ref{eq6}).

Notice that the approximated value $Q^{C}(\mathbf{s}^{(t_i)},\mathbf{a}^{(t_i)})$ and the target value $y^{(t_i)}$ are both obtained from $A$ and $C$. This results in severe over-fitting. To mitigate this problem and guarantee the convergence, the actor target network $A'$ and critic target network $C'$ are used to calculate the target
\begin{equation}\label{eq11}
y^{(t_i)}=r^{(t_i)}+\gamma Q^{C'}(\mathbf{s}^{(t_i+1)},\mathbf{a}^{(t_i+1)})|_{\mathbf{a}^{(t_i+1)}=A'(\mathbf{s}^{(t_i+1)})},
\end{equation}
so as to ensure $Q^{C}(\mathbf{s}^{(t_i)},\mathbf{a}^{(t_i)})$ and $y^{(t_i)}$ are independent. Here, $Q^{C'}(\mathbf{s}^{(t_i+1)},\mathbf{a}^{(t_i+1)})$ and $A'(\mathbf{s}^{({t_i}+1)})$ are the outputs of $C'$ and $A'$, $A'$ and $C'$ are parameterized by $\bm{\theta}^{A'}$ and $\bm{\theta}^{C'}$, respectively. The networks $A$ and $C$ are renamed as the actor and critic evaluate networks. At every learning iteration, the agent updates $A$ and $C$ by gradient descent, and the target networks $A'$ and $C'$ are soft updated by $\bm{\theta}'=\tau\bm{\theta}+(1-\tau)\bm{\theta}'$, where $\tau\ll 1$, $\bm{\theta}'$ and $\bm{\theta}$ denote the parameters of the target networks and corresponding evaluate networks respectively.

In this way, the agent can use the samples previously stored to learn an optimal beamforming policy, which improves the learning efficiency while reducing the time consumption compared with traditional approaches \cite{el2014spatially,sohrabi2016hybrid,yu2016alternating}. The steps of the proposed algorithm are summarized in Algorithm \ref{hotbooting}. Since it directly maximize $\bar{R}$ and a single agent is used to conjointly optimize the precoder and combiner, the proposed method can achieve better performance than the sub-optimal solutions optimizing the precoder and combiner separately, as shown in simulations.
\begin{algorithm}[ht]\label{algorithm1}
\centering
\caption{PrecorderNet for hybrid beamforming design}\label{hotbooting}
\begin{algorithmic}[1]
\STATE Initialize the beamforming agent; \\
       Initialize the replay buffer $D$ for the agent;\\
       Initialize the $\mathbf{s}^{(0)}$ randomly and calculate $\mathbf{V}_{RF}$;
\FOR {$t=1, 2, 3, ..., \emph{T}$}
\STATE Agent outputs $\mathbf{a}^{(t)}=A(\mathbf{s}^{(t)})+\mathbf{n}^{(t)}$;
\STATE Reorganize $\mathbf{a}^{(t)}$ into $\mathbf{V}_{BB}^{(t)}$ and $\mathbf{W}_{RF}^{(t)}$;
\STATE Normalize each element of $\mathbf{W}_{RF}^{(t)}$;
\STATE Calculate $\mathbf{W}_{BB}^{(t)}$ via (\ref{eq7});
\STATE Obtain reward $r^{(t)}$ via (\ref{eq_avgR});
\STATE Obtain $\mathbf{s}^{(t+1)}$ via (\ref{eq8});
\STATE Put transition $(\mathbf{s}^{(t)},\mathbf{a}^{(t)},r^{(t)},\mathbf{s}^{(t+1)})$ in $D$;
\STATE Sample \emph{N}-size transitions from $D$;
\STATE Update the critic evaluate network $C$ to minimize (\ref{eq10});
\STATE Update the actor evaluate network $A$ to maximize $Q^C(\mathbf{s}^{(t)},\mathbf{a}^{(t)})$;
\STATE Update the target networks by $\bm{\theta}'=\tau\bm{\theta}+(1-\tau)\bm{\theta}';$
\ENDFOR
\end{algorithmic}
\end{algorithm}

\section{Simulations}
\begin{figure}[!ht]\centering
\includegraphics[width=4.5in]{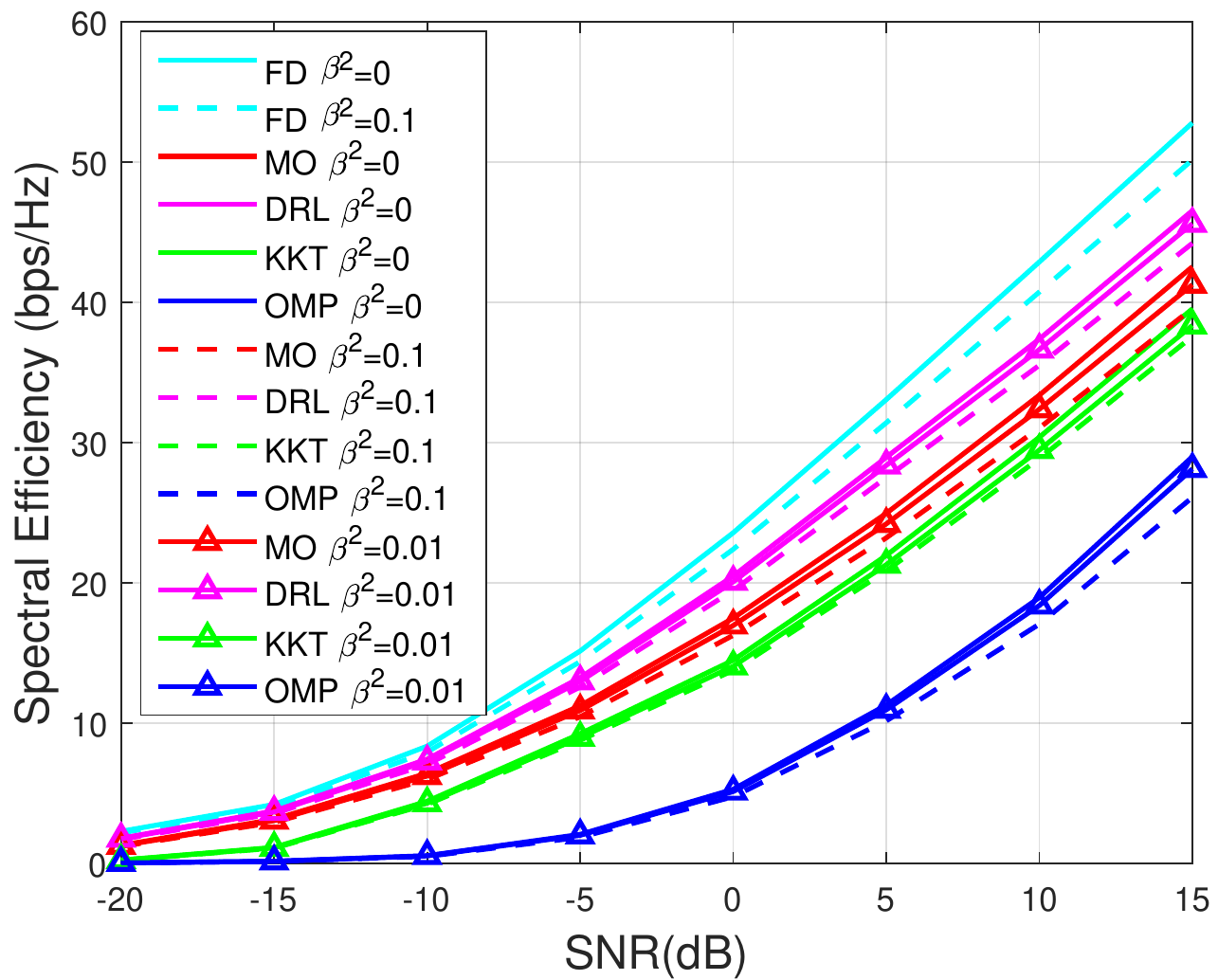}
\caption{Comparison of spectral efficiency achieved by different algorithms.}\label{fig_spectral_efficiency}
\end{figure}

\begin{figure}[!ht]\centering
\includegraphics[width=4.5in]{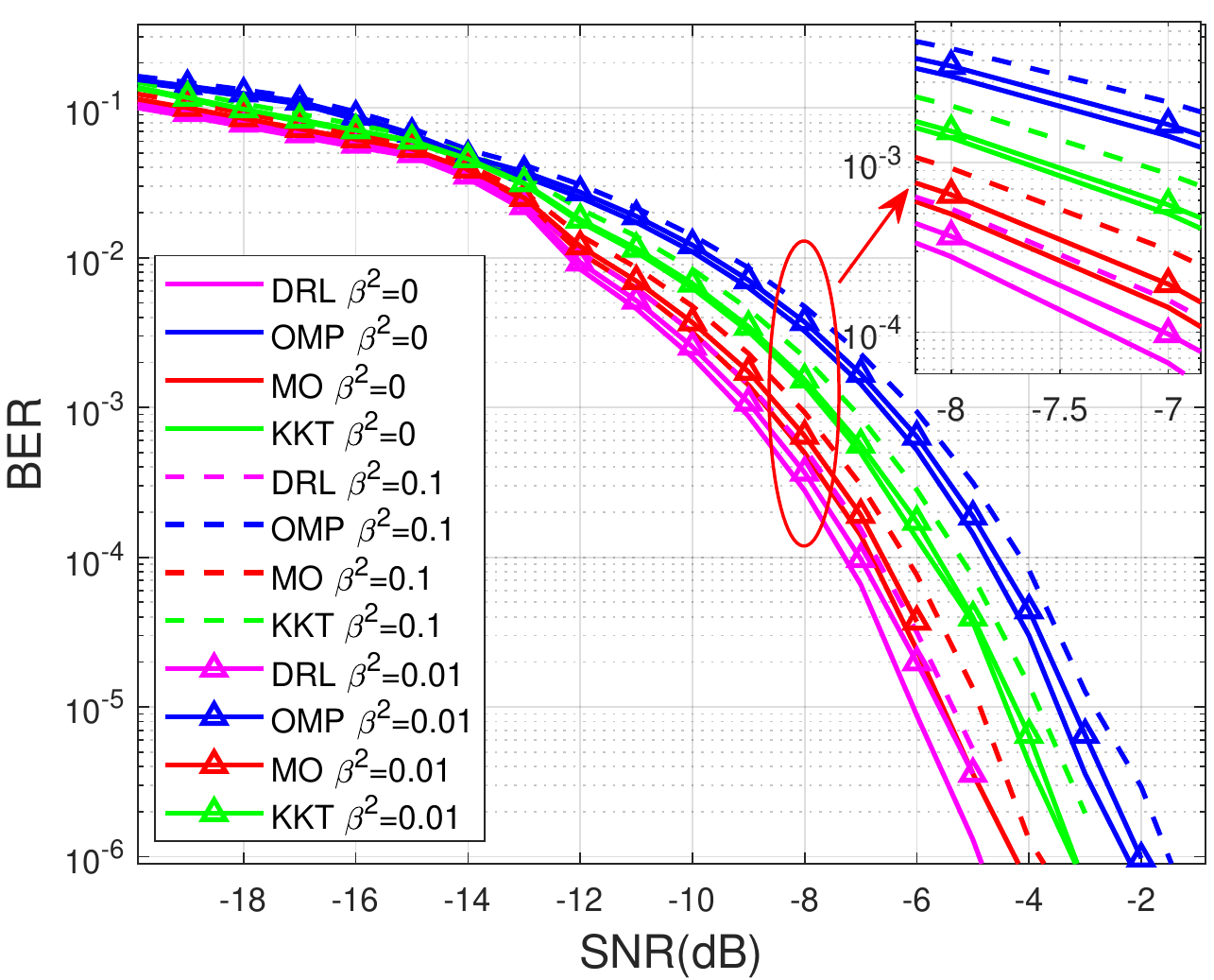}
\caption{Bit error rate of different algorithms.}\label{fig_ber}
\end{figure}

In this section, we validate the performance of the proposed algorithm. Without loss of generality, we set $d=\lambda/2$, and the threshold value of MO algorithm to calculate $\mathbf{V}_{RF}$ is set to $10^{-2}$. We first compare the performance of the proposed DRL-based algorithm with three benchmark algorithms, i.e., the OMP algorithm \cite{el2014spatially}, the KKT-based algorithm \cite{sohrabi2016hybrid}, and the MO-based algorithm \cite{yu2016alternating}, in Fig. \ref{fig_spectral_efficiency} and Fig. \ref{fig_ber}. In both figures, $N_t=128$, $N_r=32$, $N_{RF}^{t}=N_{RF}^{r}=N_s=6$, $N_{cl}=8$, $N_{ray}=10$, and  $\sigma_{\alpha,i}^2=1, \forall i$. The PrecoderNet is constructed via four-layered forward neutral networks (FNN) using Adam optimizer to perform gradient descent. Both the input and output layer have $K=2(N_{RF}^tN_s+N_rN_{RF}^r)$ neurons. The number of neuron of the two hidden layers are $300$ and $200$. The first three layers all use ReLU function as activation function. The learning rate $\alpha=0.001$, $\gamma=0.95$, $\tau=0.001$, $N=64$, $N_D=5000$, and $\mathbf{n}^{(t)}\sim\mathcal{N}(0,0.1\mathbf{I}_K)$. %The performance of the OMP algorithm proposed in \cite{el2014spatially}, the KKT-based algorithm in \cite{sohrabi2016hybrid}, and the full-digital algorithm in \cite{sampath2001generalized}, referred to as ``MMSE'' algorithm, are compared in the simulation. The ``MMSE'' algorithm employs $\mathbf{V}_{opt}$ as precoder, and obtains the combiner by MMSE criterion, which is considered as the performance upper bound in \cite{el2014spatially}. The MO-based algorithm proposed in \cite{yu2016alternating} referred to as ``MO'' is also compared.

Fig. \ref{fig_spectral_efficiency} compares the average rate achieved by the proposed PrecoderNet with the ``OMP'', ``KKT'', and ``MO'' algorithms. For the fairness of comparison, the MMSE combining matrices applied in these algorithms are also modified similarly to (\ref{eq7}) based on the level of CSI imperfection. The spectral efficiency of the full-digital algorithm that employs the right and left singular matrices of $\tilde{\mathbf{H}}$ as precoder and combiner, referred to as ``FD'', is also shown. The SNR is defined as SNR$=\frac{P}{\sigma_n^2}$, and the performance under $\beta^2=0,0.1,0.01$, corresponding to the perfect CSI, channel estimation error of -10 dB and -20 dB, are compared. It can be seen that the proposed algorithm achieves higher spectral efficiency than all the three benchmark algorithms under both perfect and imperfect CSI, especially the OMP algorithm, and is very close to the full digital algorithm. In addition, the gaps between the PrecorderNet and the benchmark algorithms increase as the SNR increases.
%Moreover, from \cite{he2018deep}, it can be seen that the channel estimation error could be as low as -20 dB to -30 dB. Under this CSI imperfection level, the proposed algorithm could provide a performance almost as good as the perfect CSI case.

Fig. \ref{fig_ber} compares the BER performance of the proposed algorithm with the ``OMP'', ``KKT'', and ``MO'' algorithms.
The BS transmits $N_{sym}=10N_{t}$ symbols per data stream and uses the quadrature phase shifting keying (QPSK) to modulate the data. Simulation results indicate that the proposed algorithm achieves much better BER performance compared with these benchmarks under both perfect and imperfect CSI. Moreover, the gaps between the proposed algorithm and the other algorithms get large as the SNR increases.

We also compare the running time of the four compared algorithms averaged over $2000$ channel realizations in Table \ref{table2}, where $\beta^2=0.01$ (time consumption of $\beta^2=0.1$ is nearly the same as $\beta^2 = 0.01$) and the other settings are the same with Fig. \ref{fig_spectral_efficiency} and Fig. \ref{fig_ber}. It can be seen that the time consumption of the PrecoderNet\footnote{The total time consumption of the PrecoderNet contains the time consumption of the MO-based approach to obtain $\mathbf{V}_{RF}$, which is about 10 ms, and the average time consumption of one learning iteration is around 0.6 ms.} is much less than the other algorithms, especially the KKT-based algorithm. Moreover, the time consumed by the OMP algorithm increases rapidly as the SNR increases, while it does not change much for the PrecoderNet. From the above results, it can be seen that proposed algorithm can achieve the best spectral efficiency and BER performance with the lowest time consumption.
%\begin{table}[ht]
%%\captionsetup{justification=centering}
%\setlength{\belowcaptionskip}{5pt}\caption{Running time comparison}\label{table2}
%\centering
%\setlength{\tabcolsep}{1mm}{
%\begin{tabular}{@{}|c|c|c|c|c|c|@{}}
%\hline
%\multirow{2}{*}{SNR (dB)} & \multicolumn{4}{c|}{Time consumption (ms)} \\ \cline{2-5}
%                  & PrecoderNet   & OMP    & KKT     & MO  \\ \hline
%-20               & 32.31         & 46.53  & 65070   & 2805   \\ \hline
%-10               & 31.56         & 79.07  & 60135   & 2792   \\ \hline
%0                 & 31.33         & 102.6  & 58024   & 2815   \\ \hline
%10                & 30.86         & 148.7  & 59195   & 2855   \\ \hline
%20                & 30.97         & 194.6  & 56326   & 2901   \\ \hline
%\end{tabular}}
%\end{table}
\begin{table}[ht]
%\captionsetup{justification=centering}
\setlength{\belowcaptionskip}{5pt}\caption{Running time comparison}\label{table2}
\centering
\setlength{\tabcolsep}{1mm}{
\begin{tabular}{@{}|c|c|c|c|c|c|@{}}
\hline
\multirow{2}{*}{SNR (dB)} & \multicolumn{4}{c|}{Time consumption (ms)} \\ \cline{2-5}
                  & PrecoderNet   & OMP    & KKT     & MO  \\ \hline
-20               & 32.52         & 48.25  & 65252   & 2915   \\ \hline
-10               & 32.50         & 80.05  & 61335   & 2901   \\ \hline
0                 & 31.64         & 105.2  & 58850   & 2877   \\ \hline
10                & 30.77         & 150.3  & 60223   & 2925   \\ \hline
20                & 30.75         & 198.6  & 57952   & 2876   \\ \hline
\end{tabular}}
\end{table}

\begin{figure}[!ht]\centering
\includegraphics[width=4.5in]{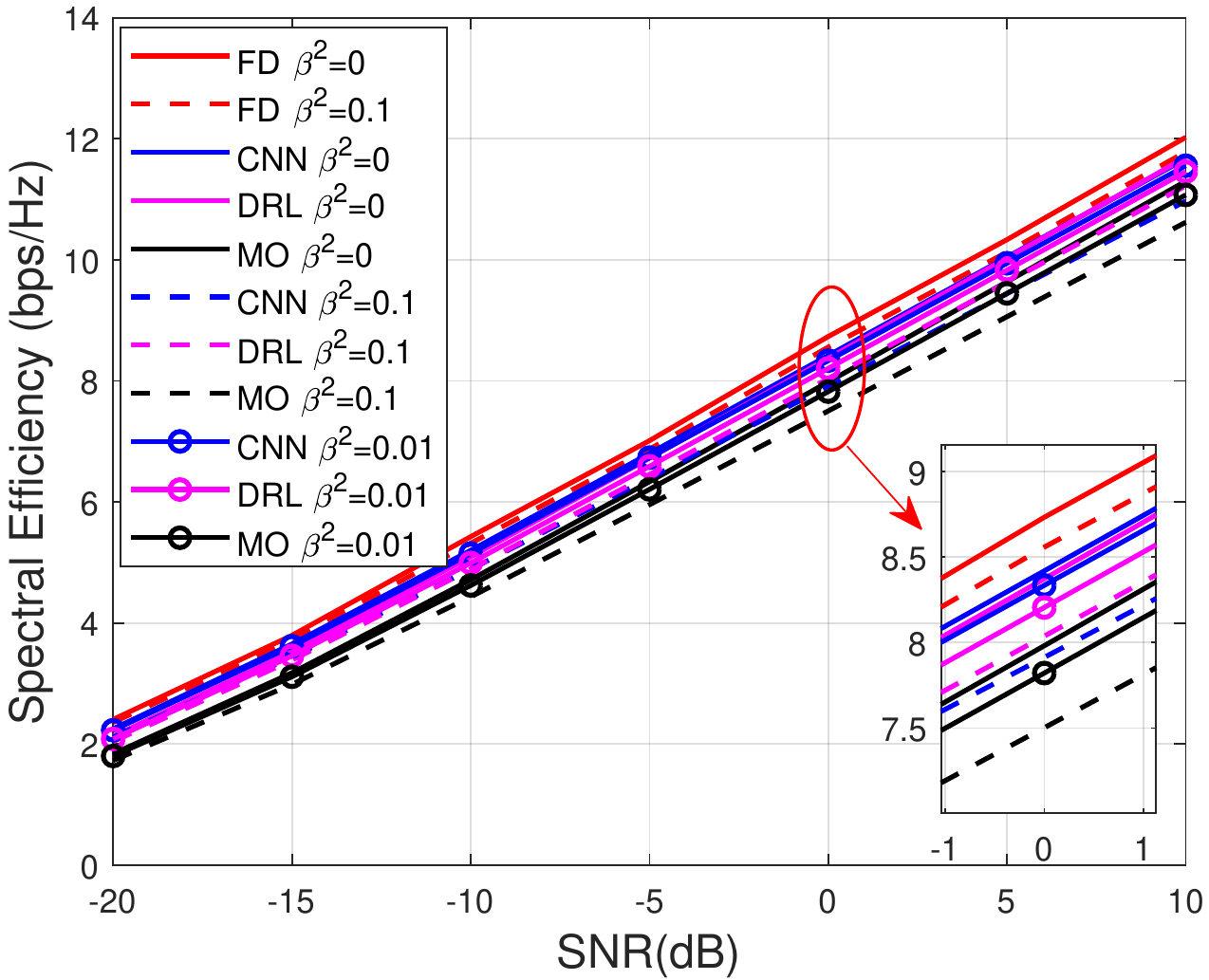}
\caption{Spectral efficiency comparison when $N_c=4$.}\label{fig_nc4}
\end{figure}

\begin{figure}[!ht]\centering
\includegraphics[width=4.5in]{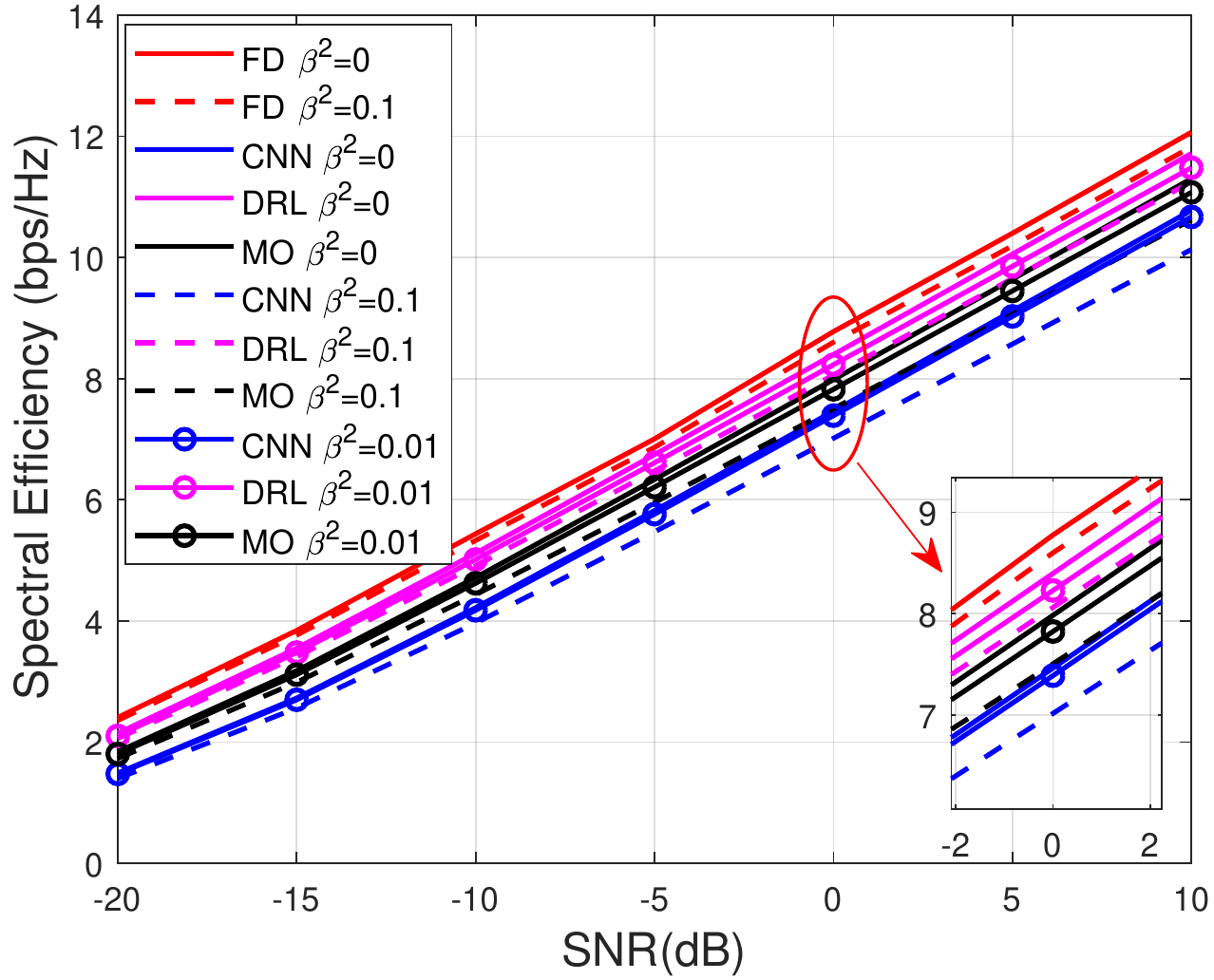}
\caption{Spectral efficiency comparison when $N_c=6$.}\label{fig_nc6}
\end{figure}

In Fig. \ref{fig_nc4} and Fig. \ref{fig_nc6}, the spectral efficiency performance and the robustness of the proposed PrecoderNet and the DL-based approach in \cite{elbir2019cnn} are compared. In Fig. \ref{fig_nc4}, we set $N_r=N_t=36$, $N_{RF}^{t}=N_{RF}^{r}=4$, $N_c=4$, and $N_{ray}=5$, as described in \cite{elbir2019cnn}. The compared convolutional neural network (CNN) model consists of two CNNs with $8$ layers which has identical structure as in \cite{elbir2019cnn}, and it is trained using the training data with exactly the same settings of the channel. The training data also takes into account three levels of channel imperfection, i.e., $\beta^2 \in \{-5,-7.5,-10 \}\mathrm{dB}$, which is the same as \cite{elbir2019cnn}. In this figure, the structure of PrecoderNet is almost the same with Fig. \ref{fig_spectral_efficiency}, except that the neuron number of the two hidden layers are $150$ and $100$. As shown in Fig. \ref{fig_nc4}, the spectral efficiency of the CNN-based and the DRL-based algorithm is nearly the same, while the performance of the CNN-based algorithm is slightly superior when $\beta^2=0$ and $0.01$. The performance of the CNN-based and the DRL-based algorithm are both very close to the full digital algorithm. For the time consumption, the PrecoderNet spends about $22.57 \mathrm{~ms}$ to compute all precoders and combiners whereas the CNN-based approach takes about $26.15 \mathrm{~ms}$.

To compare the robustness of the DL-based algorithm and the proposed DRL-based algorithm, we change the parameter $N_c$ from $4$ to $6$ while the other parameters remain unchanged. The DL-based algorithm is trained using the training data of Fig. \ref{fig_nc4}. This simulates the case when the transmission environment changes gradually. From Fig. \ref{fig_nc6}, it can be seen that the DL-based algorithm suffers a performance degradation without re-trainting, which requires new training data corresponding to the changed environment. Meanwhile, the DRL-based algorithm can adapt to the changed environment automatically, even though the CSI is imperfect. Moreover, although the on-line time consumption of the DL-based and DRL-based algorithm are very close, as shown previously, it needs at least one hour to retrain the network in the DL-based algorithm if one wants to retain the performance while the transmission environment changes. Furthermore, as the number of transmit and receive antenna, the number of data stream, and etc grow large, it becomes more and more difficult to obtain the training label for the DL-based algorithm. Thus, it can be seen that the proposed DRL-based hybrid beamforming algorithm is more effective and robust.

\section{Conclusion}
In this letter, we investigated the hybrid beamforming design for mmWave massive MIMO system, and proposed a DRL-based algorithm called PrecoderNet under imperfect CSI. Taking the precoding and combining matrices at previous learning iteration as state, while these matrices at current learning iteration as action, the DRL agent can rapidly learn the near-optimal HBF design policy under both perfect and imperfect CSI. Simulations show that the proposed algorithm performs well in spectral efficiency, BER, and time consumption.

%\clearpage
\bibliography{reference}% 指向同一个目录下的PU.bib 文件
\bibliographystyle{ieeetr}
%\end{spacing}
\end{document}